\documentclass[prl,twocolumn,amsmath,notitlepage]{revtex4-1}

\usepackage{graphicx}
\usepackage{multirow}
\usepackage{amssymb,amsfonts,amsmath}

\def\s{\sigma}

    \def \a{\alpha}
     
\def \s{\sigma}      
\def \e{\epsilon}    
   \def \d{\delta} 
    \def \l{\lambda}

\def \del{\partial}    

\def \>{\rangle} 
\def \<{\langle} 
 
\def\be{\begin{equation}} 
\def\ee{\end{equation}} 
\def\longrightharpoonup{\relbar\joinrel\rightharpoonup}
\def\longleftharpoondown{\leftharpoondown\joinrel\relbar}

\def\longrightleftharpoons{
  \mathop{
    \vcenter{
      \hbox{
      \ooalign{
        \raise1pt\hbox{$\longrightharpoonup\joinrel$}\crcr
	  \lower1pt\hbox{$\longleftharpoondown\joinrel$}
	  }
      }
    }
  }
}

\newcommand \bea {\begin{eqnarray}} 
\newcommand \eea {\end{eqnarray}}

\begin{document}

\title{Variational Pseudolikelihood for Regularized Ising Inference}

\author{Charles K. Fisher}
\affiliation{Deptartment of Physics, Boston University, Boston, MA 02215}
\email{charleskennethfisher@gmail.com}

\begin{abstract}

I propose a variational approach to maximum pseudolikelihood inference of the Ising model. The variational algorithm is more computationally efficient, and does a better job predicting out-of-sample correlations than $L_2$ regularized maximum pseudolikelihood inference as well as mean field and isolated spin pair approximations with pseudocount regularization. The key to the approach is a variational energy that regularizes the inference problem by shrinking the couplings towards zero, while still allowing some large couplings to explain strong correlations. The utility of the variational pseudolikelihood approach is illustrated by training an Ising model to represent the letters A-J using samples of letters from different computer fonts. 

 \end{abstract}

\maketitle

Statistical mechanical models constructed from experimental observations provide a valuable tool for studying complex systems. The utility of the statistical mechanics approach to data-driven modeling is especially apparent in biology, providing insights into the behavior of flocking birds \cite{bialek2012statistical}, the organization of neural networks in the brain \cite{tkavcik2013simplest,cocco2011adaptive2,roudi2009ising}, the structure and evolution of proteins \cite{cocco2013principal,ekeberg2013improved,weigt2009identification,hopf2012three,schug2009high}, and many other topics \cite{mora2010maximum,santolini2013beyond}. Generally, the `inverse' statistical mechanics approach refers to the construction of statistical models using the principle of maximum entropy \cite{jaynes1982rationale,presse2013principles}. In this approach, one constructs the probability distribution with the maximum entropy subject to constraints on its moments, which are derived from observations. Although maximum entropy modeling has been applied successfully to many different problems, it is still not clear how to estimate the parameters of the resulting probability distribution in an optimal way. 

In this work, I consider the problem of estimating the parameters of an Ising model (see \cite{roudi2009statistical} for a review). The Ising model describes the statistics of a vector $\vec{\s}$ of $N$ spin variables $\s_i \in \{-1,+1\}$, and can be derived from the principle of maximum entropy with constraints on the moments $\< \s_i \>$ and $\<\s_i \s_j\>$.  The probability distribution for $\vec{\s}$ is given by $P(\vec{\s}) = Z^{-1} \exp( - U(\vec{\s}) )$ where the energy is $U(\vec{\s}) = - \sum_i h_i \s_i - \sum_{i<j} J_{ij} \s_i \s_j $ and $Z$ is a normalization constant. Here, $h_i$ is a local field that biases $\s_i$ towards the $+1$ or $-1$ configuration, and $J_{ij}$ is a coupling that describes the strength of the interaction between spins $i$ and $j$. The goal of the inverse Ising problem is to infer the parameters, i.e.\ $h_i$ and $J_{ij}$, from a set of $n$ observed configurations $\vec{\s}^{(l)}$ for $l = 1\ldots n$. 

The Ising model presents two obstacles that make the inverse problem quite difficult. First, computing the partition function (or its derivatives) is computationally intractable for large spin systems. As a result, the inverse Ising problem has to be solved approximately.  A number of approximate methods for inferring the parameters of the Ising model have been introduced including naive mean field theory \cite{kappen1998efficient}, the Thouless-Anderson-Palmer (TAP) approximation \cite{tanaka1998mean}, the isolated spin pair approximation \cite{roudi2009statistical}, the Sessak-Monasson expansion \cite{sessak2009small}, and others \cite{aurell2012inverse,nguyen2012bethe,cocco2013principal,mastromatteo2013beyond,huang2013sparse,cocco2012adaptive,nguyen2012mean,huang2013adaptive,
cocco2011adaptive,huang2010reconstructing,aurell2010dynamics}. The second obstacle -- overfitting -- is more fundamental and generally affects all high dimensional problems in statistical inference. 

Overfitting is a simple concept to illustrate for the Ising model.  All of the moments, i.e.\ $\<\s_i\>$ and $\<\s_i \s_j\>$, are noisy because they are computed from a finite sample of $n$ observed configurations. Thus, fitting all of the moments exactly necessarily incorporates noise from the finite sample size. As a consequence, parameters obtained by fitting the moments of one dataset may not provide a good description of a new dataset derived from the same system. In general, overfitting becomes a serious problem when the number of parameters (i.e.\ $N^2$) exceeds the number of independent observations (i.e. $n$). 

A common approach to mitigating the effects of overfitting is to penalize parameters with large values \cite{barton2014large}. For example, the Ising model can be `regularized' by adding an $L_2$ penalty $\lambda \sum_{i<j} J_{ij}^2$ (as in \cite{barton2014large}) or an $L_1$ penalty $\lambda \sum_{i<j} |J_{ij}|$ (as in \cite{aurell2012inverse,huang2013sparse}) to the objective function that describes the fit to the data. Alternatively, the empirical moments can be modified using a `pseudocount' according to the rules $\<\s_i\> \to (1- \a) \< \s_i\>$ and $\<\s_i \s_j \> \to \d_{ij} + (1-\d_{ij}) (1-\a) \<\s_i \s_j \>$ \cite{barton2014large}. The intuition for why regularization works comes from the law of total variance, which implies that the error in a statistical estimator is $error^2 = variance + bias^2$. Thus, a regularization method that decreases the variance in an estimator more than it increases the squared bias will have a smaller error and, therefore, better out-of-sample predictive ability. 

The free parameter ($\l \geq 0$ or $0 \leq \a \leq 1$) that penalizes large couplings must be carefully chosen to ensure that the regularization method actually improves out-of-sample performance. Typically, the choice of the regularization parameter and subsequent testing of predictive ability are performed using cross-validation. To perform cross-validation, the observed data are randomly partitioned into three mutually exclusive datasets. The first dataset, usually called the `training' sample, is used for fitting the model. Here, the training sample will be denoted `F' for `fitting'. The resulting parameters are used to predict the data contained in the second dataset, called the 'validation' sample (`V'). The optimal value for the regularization parameter is chosen to maximize the predictive performance of the model on the validation sample. Finally, the out-of-sample performance of the resulting model is tested by measuring the agreement between the predictions of the model and the data contained in the third dataset, called the 'test' sample (`T').

\begin{figure}[t]
\includegraphics[width=3in]{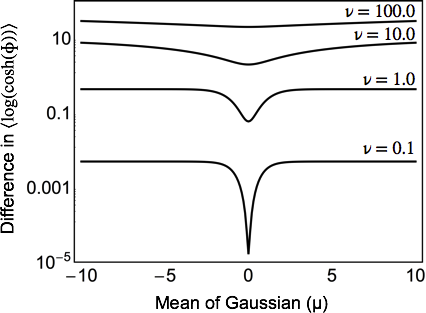}
\caption{A figure demonstrating that $ \log(\cosh(\mu)) + \log(\cosh(\s)) - \< \log(\cosh(\phi)) \>$ (the vertical axis) is greater than zero when $\phi$ is Gaussian distributed with mean $\mu$ and standard deviation $\nu$. }
\label{fig:fig1}
\end{figure}

Focusing on out-of-sample predictive ability provides a different metric than most previous papers on the inverse Ising model, which have studied the ability to reconstruct a coupling matrix from sampled configurations (e.g.\ \cite{roudi2009statistical}). In general, the performance of various algorithms for the inverse Ising model have been studied by defining a set of couplings, simulating configurations from the resulting Ising model, and then asking how well the couplings inferred by various methods reproduce the true $J_{ij}$'s. By contrast, this work will compare the performance of different approaches to the inverse Ising model by their out-of-sample predictive performance using a dataset of images developed for testing machine learning algorithms. 

In order to compare the predictive performance of various inference methods for the Ising model, it is necessary to have a metric that quantifies the agreement between the model and the data in the test sample that can be easily computed.  Here, I will use the negative log-pseudolikelihood, $\mathcal{L}^{data}_{p}(h,J) = -\< \log \prod_i P(\s_i | \s_{j \neq i} ) \>_{data}$, where the angular brackets denote an average over the spin configurations in dataset `F', `V', or `T'. Recently, Aurell and Ekeberg \cite{aurell2012inverse} demonstrated that the couplings of the Ising model can be inferred by minimizing the negative log-pseudolikelihood, which is given by (ignoring constant terms):
\begin{align}
\label{eq:log-pseudo}
&\mathcal{L}^F_{p}(h,J) = - \< \log \prod_i P(\s_i | \s_{j \neq i} ) \>_{F} \\
& = - \sum_i h_i \<\s_i\>_{F} - \sum_{i, j\neq i}J_{ij} \<\s_i \s_j\>_{F} + \sum_i \< \log (\cosh( \phi_i )) \>_{F} \nonumber
\end{align}
where $ \phi_i = h_i + \sum_{j \neq i} J_{ij} \s_j$ is the effective field acting on spin $i$ due to the configurations of the other $n-1$ spins. The derivatives of $\mathcal{L}_{p}(h,J)$ are easy to calculate:
\begin{align}
\frac{\del \mathcal{L}^F_p}{\del h_i} &= \<\s_i\>_F - \< \tanh(\phi_i ) \>_F \nonumber \\
\frac{\del \mathcal{L}^F_p}{\del J_{ij}} &= 2\<\s_i \s_j \>_F - \< v_i \tanh(\phi_j ) \>_F - \< v_j \tanh(\phi_i) \>_F \nonumber
\end{align}
Note that these expressions assume $J_{ii} = 0$ and $J_{ij} = J_{ji}$. Thus, the Ising model can be fit to the training sample by minimizing $\mathcal{L}^F_{p}(h,J)$ using gradient descent. 

\begin{figure}[b]
\includegraphics[width=3in]{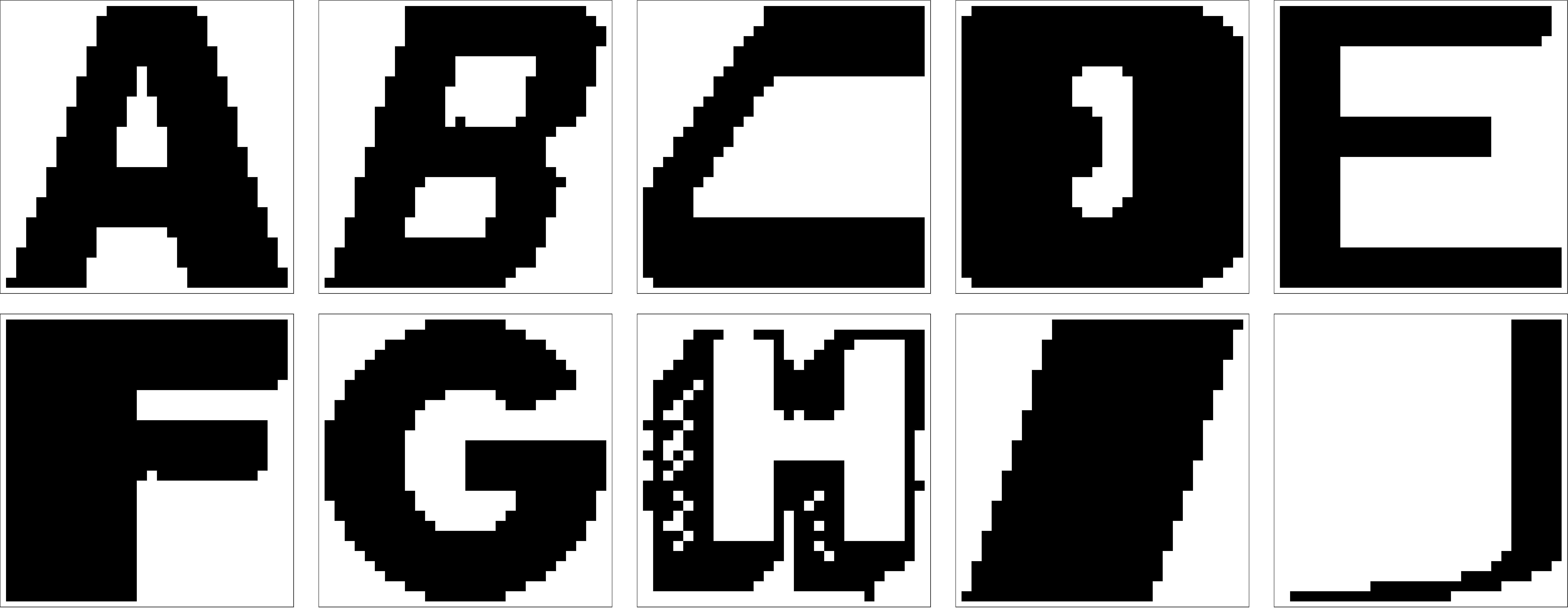}
\caption{Randomly chosen black and white images of letters A-J from the letters dataset.}
\label{fig:fig2}
\end{figure}

While the computing the pseudolikelihood is certainly tractable it is, neverthless, computationally challenging to compute the averages over the training sample at every step of gradient descent if the sample size is large. Therefore, I propose a variational approach to maximum pseudolikelihood inference of the Ising model that is more computationally efficient and, as I will show, performs better out of sample. The variational pseudolikelihood approach minimizes an energy that provides an upper bound on $\mathcal{L}^F_{p}(h,J)$, given by:
\begin{align}
\label{eq:variational-energy}
&\mathcal{E}(h,J) = - \sum_i h_i \<\s_i\>_{F} - \sum_{i} \sum_{j \neq i} J_{ij} \<\s_i \s_j\>_{F} \nonumber \\
&+ \sum_i ( \log (\cosh( \mu_i )) + \log(\cosh(\nu_i) ) ) 
\end{align}
The upper bound follows from $\< \log(\cosh(\phi_i)) \>  \leq \log(\cosh(\mu_i)) + \log(\cosh(\nu_i))$ if $\phi_i$ is Gaussian distributed with mean $\mu_i$ and standard deviation $\nu_i$; this inequality is illustrated in Fig. \ref{fig:fig1}. Since $\phi_i$ is a sum of many random variables, I will assume that it is approximately Gaussian distributed with mean $\mu_i = h_i + \sum_{j \neq i} J_{ij} m_j$ and variance $\nu_i^2 = \sum_{j \neq i} \sum_{k \neq i} J_{ij} J_{ik} C_{jk}$, where $m_i = \< s_i \>_F$, $C_{ij} = \<s_i s_j \>_F - \<s_i\>_F\<s_j\>_F$, and angular brackets denote averages over the training sample `F'. Minimizing the variational energy over $h_i$ yields the estimate $\hat{h}_i = \tanh^{-1}(m_i) - \sum_{j \neq i} J_{ij} m_j$. After plugging in the estimate for $h_i$, it is only necessary to minimize a function of $J$ given by:
\begin{align}
\label{eq:variational-energy-J}
G(J) & = - \sum_{i} \sum_{j \neq i} J_{ij} C_{ij} + \sum_i  \log(\cosh(\nu_i))
\end{align}
The derivatives are given by:
\be
\frac{\del G}{\del J_{ij} } = (DJC)_{ij} + (DJC)_{ji} - 2 C_{ij} = G'_{ij}
\ee
where $D_{ij} = \tanh(\nu_i) \d_{ij} / \nu_i$ and the diagonal elements of $J$ are constrained to $J_{ii} = 0$. The couplings can be estimated by minimizing Eq.\ \ref{eq:variational-energy-J} using the gradient descent update $J_{ij}(t+1) = J_{ij}(t) - \e ( G'_{ij}(t) + \rho G'_{ij}(t-1) )$, where $\e$ is a small step size and $\rho$ is a `momentum'. 

The variational pseudolikelihood approach was compared to (\cite{roudi2009statistical}): direct pseudolikelihood maximization (with and without an $L_2$ penalty on the couplings), naive mean field (NMF) inversion (with and without pseudocount regularization), and the isolated spin pair (ISP) approximation (with and without pseudocount regularization). For the regularized methods, the penalty parameters were chosen by cross-validation to minimize $\mathcal{L}^V_p(h,J)$ on the validation sample. The performances of these methods was assessed using training, validation, and test samples con structured from the `notMNIST' dataset, which consists of $28 \times 28$ pixel images of the letters A-J compiled from a large number of different fonts (see Appendix). The `notMNIST' dataset contains roughly 50000 images for each letter, and the samples of each letter are very diverse. Examples of binarized images for letters A-J are shown in Fig.\ \ref{fig:fig2}. Mutually exclusive training, validation, and test samples were constructed, each containing 500 randomly chosen images for each letter.

\begin{table}[t]
\caption{\label{tab:table1} Comparison of different methods for Ising inference according to their negative log-pseudolikelihoods on a dataset of letters. A more negative value indicates a better fit. The best performing methods for the training sample and the test sample are indicated by bold font. Pseudolikelihood and Variational Pseudolikelihood were fit using gradient descent with $\e = 0.01$ and $\rho = 0.5$ for $10000$ steps. In all cases, the local fields were estimated using $\hat{h}_i = \tanh^{-1}(m_i) - \sum_{j \neq i} J_{ij} m_j$.}
\begin{ruledtabular}
\begin{tabular}{|l|r|r|}
Method & Training Sample & Test Sample \\
\hline
Variational Pseudolikelihood & - 475.2 & \bf{- 445.7} \\
\hline
Pseudolikelihood & \bf{-535.8} & -303.3 \\
\hline
Pseudolikelihood + $L_2$ & -414.3 & -398.9 \\
\hline
Naive Mean Field (NMF) & -443.1 & -282.5 \\
\hline
NMF with pseudocount & - 423.8 & -310.4 \\
\hline
Isolated Spin Pair (ISP) & 44.7 & 46.5 \\
\hline
ISP with pseudocount & -109.2 & -108.1 \\
\end{tabular}
\end{ruledtabular}
\label{table:table1}
\end{table}

\begin{figure}[b]
\includegraphics[width=3in]{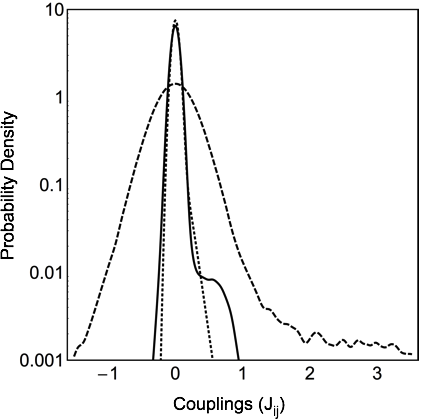}
\caption{Smooth histograms of the inferred couplings obtained from direct pseudolikelihood maximization (dashed line), pseudolikelihood maximization with an $L_2$ penalty (dotted line), and the variational pseudolikelihood method (solid line).}
\label{fig:fig3}
\end{figure}

Table \ref{table:table1} shows a comparision of these methods according to $\mathcal{L}^F_p(h,J)$ for the training set images, and $\mathcal{L}^T_p(h,J)$ for the test set images. As one would expect, ranking by the pseudolikelihood ensures that direct pseudolikelihood maximization is the best performing method on the training data. However, the variational pseudolikelihood approach presented here outperforms all other methods, including direct pseudolikelihood maximization, when compared by out-of-sample performance on the test set. In addition, a simple implementation of variational pseudolikelihood inference was approximately an order of magnitude faster than direct pseudolikelihood maximization running on a desktop computer, even ignoring the time spent choosing the penalty parameter for $L_2$ regularized pseudolikelihood. 

\begin{figure*}[t]
\includegraphics[width=7.0in]{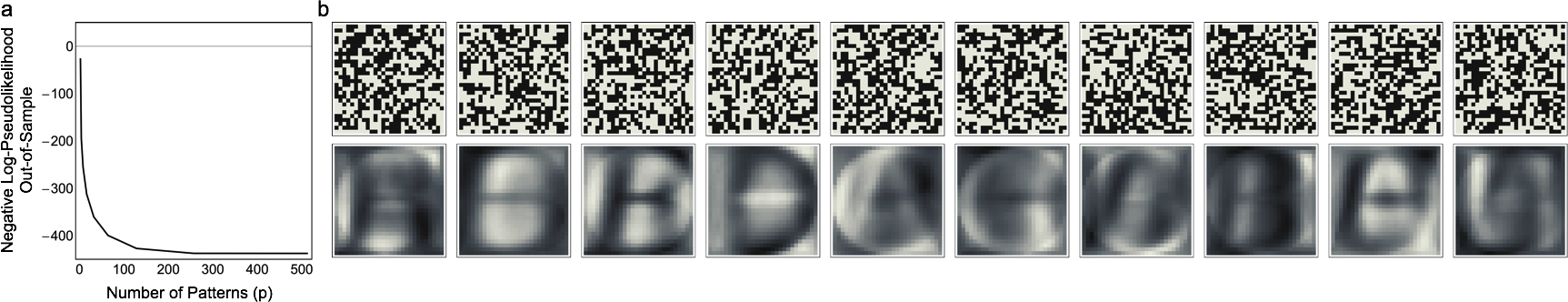}
\caption{Inference of Hopfield-like patterns using variational pseudolikelihood. a) Negative log-pseudolikelihood on the test sample as a function of the number of patterns. b) The top row shows the random initial conditions for the 10 patterns. The bottom row shows the 10 patterns obtained by gradient descent minimization of the variational pseudolikelihood with $\e = 0.001$ and $\rho = 0.5$. The model achieved negative log-pseudolikelihoods of $\mathcal{L}^F_p(h,J) = -277.1$ and $\mathcal{L}^T_p(h,J) = -275.0$ on the training sample and the test sample, respectively. }
\label{fig:fig4}
\end{figure*} 

Comparing the out-of-sample results for the unregularized and regularized versions of maximum pseudolikelihood, NMF, and ISP estimates in Table \ref{table:table1} demonstrates the importance of regularization for the inverse Ising model. Histograms of the couplings inferred with direct pseudolikelihood maximization, pseudlikelihood maximization with an $L_2$ penalty, and the variational pseudolikelihood method (Fig.\ \ref{fig:fig2}) illustrate that the methods which performed better out-of-sample had smaller estimates for the couplings. However, adding the $L_2$ penalty to the pseudolikelihood resulted in an inferred Ising model with many weak couplings, and no strong couplings. The variational approach, by contrast, produced both strong and weak couplings, which resulted in improved out-of-sample prediction performance relative to other methods. 

The analysis presented above demonstrates that variational pseudolikelihood inference can be used to train an Ising model so that it makes good out-of-sample predictions. Here, I provide an example to illustrate how the variational pseudolikelihood method can be used in practice. I examined a simple question: what features does the Ising model learn when trained on images of the letters A-J? Using the variational pseudolikelihood approach, it is simple to infer coupling matrices with a specific structure, such as that of a Hopfield neural network \cite{hopfield1982neural,huang2010reconstructing,huang2013sparse,cocco2013principal}. The couplings in a Hopfield network are described by a Hebbian rule such that $J_{ij} = \sum_{s} \xi^{(s)}_i \xi^{(s)}_j (1- \d_{ij})$, where $\vec{\xi}^{(s)}$ for $s = 1 \ldots p$ are `patterns' stored in the memory. These patterns describe features in the correlation structure of the data. The derivatives of the variational energy with respect to the patterns can be computed using the chain rule:
\be
\frac{ \del G}{\del \xi^{(s)}_i} \propto \sum_j G'_{ij} \xi^{(s)}_j
\ee
Thus, the patterns can be inferred quite easily by minimizing the variational energy using gradient descent. Note, however, that the patterns are unconstrained with respect to orthogonal transformations \cite{cocco2013inference}, and the symmetry is broken by the initial conditions of the optimization. 

The letters in the notMNIST dataset are quiet diverse, and hundreds of patterns are required to approach the performance of the unconstrained variational pseudolikelihood method (Fig.\ \ref{fig:fig4}a). Nevertheless, the $p = 10$ Hopfield model provides a useful visual example. Starting from randomly initialized patterns (shown in the top row of Fig.\ \ref{fig:fig4}b), the patterns converge after less than $1000$ steps of gradient descent to images with recognizable features of letters (shown in the second row of Fig.\ \ref{fig:fig4}b). 

The notMNIST images present a challenging dataset for the Ising model. State-of-the-art machine learning methods for modeling similar types of images include a variety of pre-processing and fine tuning steps \cite{tang2011data}, and the best performing algorithms are based on deep neural networks that are believed to excel at capturing higher order correlation structure in the data that cannot be described by the Ising model \cite{ciresan2012multi}. Nevertheless, Fig.\ \ref{fig:fig4} clearly demonstrates an Ising model based on a few Hopfield patterns is able to capture recognizable features from the data. 

In summary, I presented a variational approach to maximum pseudolikelihood inference of the Ising model. The variational pseudolikelihood method outperformed all other tested methods at describing the out-of-sample correlation structure in a dataset consisting of images of the letters A-J from various computer fonts. This method has many characteristics that make it attractive in practice: it is fast, accurate, can be used with large spin systems, and can easily be adapted to learn coupling matrices with specific structures such as Hopfield models. Moreover, it should be easy to extend the approach presented here to construct variational pseudolikelihood algorithms for inferring Potts models \cite{ekeberg2013improved}, Restricted Boltzmann Machines \cite{salakhutdinov2007restricted}, and other inverse models for spin systems. 

\begin{acknowledgments}
I would like to thank Alex H. Lang, Thomas C. Lang, Pankaj Mehta, Ilya Nemenman, and Javad Noorbakhsh for helpful discussions. Funding for this research was provided by a Simons Investigator Award to Pankaj Mehta. 
\end{acknowledgments}

\appendix{Appendix:}
The `notMNIST' dataset consists of $28 \times 28$ pixel images of the letters A-J compiled from a large number of different fonts. The images are available at http://yaroslavvb.blogspot.com/2011/09/notmnist-dataset.html. Each image in the original dataset is grayscale, with pixels taking on values between $0$ and $255$. The images were binarized in the simplest way by defining pixels with intensity less than or equal to $127$ as spin $-1$, and those with intensity greater than $127$ as spin $+1$. Finally, each image was flattened into a vector of $N = 784$ spins. 

\bibliography{variational_refs}

\begin{thebibliography}{33}%
\makeatletter
\providecommand \@ifxundefined [1]{%
 \@ifx{#1\undefined}
}%
\providecommand \@ifnum [1]{%
 \ifnum #1\expandafter \@firstoftwo
 \else \expandafter \@secondoftwo
 \fi
}%
\providecommand \@ifx [1]{%
 \ifx #1\expandafter \@firstoftwo
 \else \expandafter \@secondoftwo
 \fi
}%
\providecommand \natexlab [1]{#1}%
\providecommand \enquote  [1]{``#1''}%
\providecommand \bibnamefont  [1]{#1}%
\providecommand \bibfnamefont [1]{#1}%
\providecommand \citenamefont [1]{#1}%
\providecommand \href@noop [0]{\@secondoftwo}%
\providecommand \href [0]{\begingroup \@sanitize@url \@href}%
\providecommand \@href[1]{\@@startlink{#1}\@@href}%
\providecommand \@@href[1]{\endgroup#1\@@endlink}%
\providecommand \@sanitize@url [0]{\catcode `\\12\catcode `\$12\catcode
  `\&12\catcode `\#12\catcode `\^12\catcode `\_12\catcode `\%12\relax}%
\providecommand \@@startlink[1]{}%
\providecommand \@@endlink[0]{}%
\providecommand \url  [0]{\begingroup\@sanitize@url \@url }%
\providecommand \@url [1]{\endgroup\@href {#1}{\urlprefix }}%
\providecommand \urlprefix  [0]{URL }%
\providecommand \Eprint [0]{\href }%
\providecommand \doibase [0]{http://dx.doi.org/}%
\providecommand \selectlanguage [0]{\@gobble}%
\providecommand \bibinfo  [0]{\@secondoftwo}%
\providecommand \bibfield  [0]{\@secondoftwo}%
\providecommand \translation [1]{[#1]}%
\providecommand \BibitemOpen [0]{}%
\providecommand \bibitemStop [0]{}%
\providecommand \bibitemNoStop [0]{.\EOS\space}%
\providecommand \EOS [0]{\spacefactor3000\relax}%
\providecommand \BibitemShut  [1]{\csname bibitem#1\endcsname}%
\let\auto@bib@innerbib\@empty
\bibitem [{\citenamefont {Bialek}\ \emph {et~al.}(2012)\citenamefont {Bialek},
  \citenamefont {Cavagna}, \citenamefont {Giardina}, \citenamefont {Mora},
  \citenamefont {Silvestri}, \citenamefont {Viale},\ and\ \citenamefont
  {Walczak}}]{bialek2012statistical}%
  \BibitemOpen
  \bibfield  {author} {\bibinfo {author} {\bibfnamefont {W.}~\bibnamefont
  {Bialek}}, \bibinfo {author} {\bibfnamefont {A.}~\bibnamefont {Cavagna}},
  \bibinfo {author} {\bibfnamefont {I.}~\bibnamefont {Giardina}}, \bibinfo
  {author} {\bibfnamefont {T.}~\bibnamefont {Mora}}, \bibinfo {author}
  {\bibfnamefont {E.}~\bibnamefont {Silvestri}}, \bibinfo {author}
  {\bibfnamefont {M.}~\bibnamefont {Viale}}, \ and\ \bibinfo {author}
  {\bibfnamefont {A.~M.}\ \bibnamefont {Walczak}},\ }\href@noop {} {\bibfield
  {journal} {\bibinfo  {journal} {Proceedings of the National Academy of
  Sciences}\ }\textbf {\bibinfo {volume} {109}},\ \bibinfo {pages} {4786}
  (\bibinfo {year} {2012})}\BibitemShut {NoStop}%
\bibitem [{\citenamefont {Tka{\v{c}}ik}\ \emph {et~al.}(2013)\citenamefont
  {Tka{\v{c}}ik}, \citenamefont {Marre}, \citenamefont {Mora}, \citenamefont
  {Amodei}, \citenamefont {Berry~II},\ and\ \citenamefont
  {Bialek}}]{tkavcik2013simplest}%
  \BibitemOpen
  \bibfield  {author} {\bibinfo {author} {\bibfnamefont {G.}~\bibnamefont
  {Tka{\v{c}}ik}}, \bibinfo {author} {\bibfnamefont {O.}~\bibnamefont {Marre}},
  \bibinfo {author} {\bibfnamefont {T.}~\bibnamefont {Mora}}, \bibinfo {author}
  {\bibfnamefont {D.}~\bibnamefont {Amodei}}, \bibinfo {author} {\bibfnamefont
  {M.~J.}\ \bibnamefont {Berry~II}}, \ and\ \bibinfo {author} {\bibfnamefont
  {W.}~\bibnamefont {Bialek}},\ }\href@noop {} {\bibfield  {journal} {\bibinfo
  {journal} {Journal of Statistical Mechanics: Theory and Experiment}\ }\textbf
  {\bibinfo {volume} {2013}},\ \bibinfo {pages} {P03011} (\bibinfo {year}
  {2013})}\BibitemShut {NoStop}%
\bibitem [{\citenamefont {Cocco}\ and\ \citenamefont
  {Monasson}(2011{\natexlab{a}})}]{cocco2011adaptive2}%
  \BibitemOpen
  \bibfield  {author} {\bibinfo {author} {\bibfnamefont {S.}~\bibnamefont
  {Cocco}}\ and\ \bibinfo {author} {\bibfnamefont {R.}~\bibnamefont
  {Monasson}},\ }\href@noop {} {\bibfield  {journal} {\bibinfo  {journal} {BMC
  Neuroscience}\ }\textbf {\bibinfo {volume} {12}},\ \bibinfo {pages} {P224}
  (\bibinfo {year} {2011}{\natexlab{a}})}\BibitemShut {NoStop}%
\bibitem [{\citenamefont {Roudi}\ \emph
  {et~al.}(2009{\natexlab{a}})\citenamefont {Roudi}, \citenamefont {Tyrcha},\
  and\ \citenamefont {Hertz}}]{roudi2009ising}%
  \BibitemOpen
  \bibfield  {author} {\bibinfo {author} {\bibfnamefont {Y.}~\bibnamefont
  {Roudi}}, \bibinfo {author} {\bibfnamefont {J.}~\bibnamefont {Tyrcha}}, \
  and\ \bibinfo {author} {\bibfnamefont {J.}~\bibnamefont {Hertz}},\
  }\href@noop {} {\bibfield  {journal} {\bibinfo  {journal} {Physical Review
  E}\ }\textbf {\bibinfo {volume} {79}},\ \bibinfo {pages} {051915} (\bibinfo
  {year} {2009}{\natexlab{a}})}\BibitemShut {NoStop}%
\bibitem [{\citenamefont {Cocco}\ \emph
  {et~al.}(2013{\natexlab{a}})\citenamefont {Cocco}, \citenamefont {Monasson},\
  and\ \citenamefont {Weigt}}]{cocco2013principal}%
  \BibitemOpen
  \bibfield  {author} {\bibinfo {author} {\bibfnamefont {S.}~\bibnamefont
  {Cocco}}, \bibinfo {author} {\bibfnamefont {R.}~\bibnamefont {Monasson}}, \
  and\ \bibinfo {author} {\bibfnamefont {M.}~\bibnamefont {Weigt}},\
  }\href@noop {} {\bibfield  {journal} {\bibinfo  {journal} {PLoS computational
  biology}\ }\textbf {\bibinfo {volume} {9}},\ \bibinfo {pages} {e1003176}
  (\bibinfo {year} {2013}{\natexlab{a}})}\BibitemShut {NoStop}%
\bibitem [{\citenamefont {Ekeberg}\ \emph {et~al.}(2013)\citenamefont
  {Ekeberg}, \citenamefont {L{\"o}vkvist}, \citenamefont {Lan}, \citenamefont
  {Weigt},\ and\ \citenamefont {Aurell}}]{ekeberg2013improved}%
  \BibitemOpen
  \bibfield  {author} {\bibinfo {author} {\bibfnamefont {M.}~\bibnamefont
  {Ekeberg}}, \bibinfo {author} {\bibfnamefont {C.}~\bibnamefont
  {L{\"o}vkvist}}, \bibinfo {author} {\bibfnamefont {Y.}~\bibnamefont {Lan}},
  \bibinfo {author} {\bibfnamefont {M.}~\bibnamefont {Weigt}}, \ and\ \bibinfo
  {author} {\bibfnamefont {E.}~\bibnamefont {Aurell}},\ }\href@noop {}
  {\bibfield  {journal} {\bibinfo  {journal} {Physical Review E}\ }\textbf
  {\bibinfo {volume} {87}},\ \bibinfo {pages} {012707} (\bibinfo {year}
  {2013})}\BibitemShut {NoStop}%
\bibitem [{\citenamefont {Weigt}\ \emph {et~al.}(2009)\citenamefont {Weigt},
  \citenamefont {White}, \citenamefont {Szurmant}, \citenamefont {Hoch},\ and\
  \citenamefont {Hwa}}]{weigt2009identification}%
  \BibitemOpen
  \bibfield  {author} {\bibinfo {author} {\bibfnamefont {M.}~\bibnamefont
  {Weigt}}, \bibinfo {author} {\bibfnamefont {R.~A.}\ \bibnamefont {White}},
  \bibinfo {author} {\bibfnamefont {H.}~\bibnamefont {Szurmant}}, \bibinfo
  {author} {\bibfnamefont {J.~A.}\ \bibnamefont {Hoch}}, \ and\ \bibinfo
  {author} {\bibfnamefont {T.}~\bibnamefont {Hwa}},\ }\href@noop {} {\bibfield
  {journal} {\bibinfo  {journal} {Proceedings of the National Academy of
  Sciences}\ }\textbf {\bibinfo {volume} {106}},\ \bibinfo {pages} {67}
  (\bibinfo {year} {2009})}\BibitemShut {NoStop}%
\bibitem [{\citenamefont {Hopf}\ \emph {et~al.}(2012)\citenamefont {Hopf},
  \citenamefont {Colwell}, \citenamefont {Sheridan}, \citenamefont {Rost},
  \citenamefont {Sander},\ and\ \citenamefont {Marks}}]{hopf2012three}%
  \BibitemOpen
  \bibfield  {author} {\bibinfo {author} {\bibfnamefont {T.~A.}\ \bibnamefont
  {Hopf}}, \bibinfo {author} {\bibfnamefont {L.~J.}\ \bibnamefont {Colwell}},
  \bibinfo {author} {\bibfnamefont {R.}~\bibnamefont {Sheridan}}, \bibinfo
  {author} {\bibfnamefont {B.}~\bibnamefont {Rost}}, \bibinfo {author}
  {\bibfnamefont {C.}~\bibnamefont {Sander}}, \ and\ \bibinfo {author}
  {\bibfnamefont {D.~S.}\ \bibnamefont {Marks}},\ }\href@noop {} {\bibfield
  {journal} {\bibinfo  {journal} {Cell}\ }\textbf {\bibinfo {volume} {149}},\
  \bibinfo {pages} {1607} (\bibinfo {year} {2012})}\BibitemShut {NoStop}%
\bibitem [{\citenamefont {Schug}\ \emph {et~al.}(2009)\citenamefont {Schug},
  \citenamefont {Weigt}, \citenamefont {Onuchic}, \citenamefont {Hwa},\ and\
  \citenamefont {Szurmant}}]{schug2009high}%
  \BibitemOpen
  \bibfield  {author} {\bibinfo {author} {\bibfnamefont {A.}~\bibnamefont
  {Schug}}, \bibinfo {author} {\bibfnamefont {M.}~\bibnamefont {Weigt}},
  \bibinfo {author} {\bibfnamefont {J.~N.}\ \bibnamefont {Onuchic}}, \bibinfo
  {author} {\bibfnamefont {T.}~\bibnamefont {Hwa}}, \ and\ \bibinfo {author}
  {\bibfnamefont {H.}~\bibnamefont {Szurmant}},\ }\href@noop {} {\bibfield
  {journal} {\bibinfo  {journal} {Proceedings of the National Academy of
  Sciences}\ }\textbf {\bibinfo {volume} {106}},\ \bibinfo {pages} {22124}
  (\bibinfo {year} {2009})}\BibitemShut {NoStop}%
\bibitem [{\citenamefont {Mora}\ \emph {et~al.}(2010)\citenamefont {Mora},
  \citenamefont {Walczak}, \citenamefont {Bialek},\ and\ \citenamefont
  {Callan}}]{mora2010maximum}%
  \BibitemOpen
  \bibfield  {author} {\bibinfo {author} {\bibfnamefont {T.}~\bibnamefont
  {Mora}}, \bibinfo {author} {\bibfnamefont {A.~M.}\ \bibnamefont {Walczak}},
  \bibinfo {author} {\bibfnamefont {W.}~\bibnamefont {Bialek}}, \ and\ \bibinfo
  {author} {\bibfnamefont {C.~G.}\ \bibnamefont {Callan}},\ }\href@noop {}
  {\bibfield  {journal} {\bibinfo  {journal} {Proceedings of the National
  Academy of Sciences}\ }\textbf {\bibinfo {volume} {107}},\ \bibinfo {pages}
  {5405} (\bibinfo {year} {2010})}\BibitemShut {NoStop}%
\bibitem [{\citenamefont {Santolini}\ \emph {et~al.}(2013)\citenamefont
  {Santolini}, \citenamefont {Mora},\ and\ \citenamefont
  {Hakim}}]{santolini2013beyond}%
  \BibitemOpen
  \bibfield  {author} {\bibinfo {author} {\bibfnamefont {M.}~\bibnamefont
  {Santolini}}, \bibinfo {author} {\bibfnamefont {T.}~\bibnamefont {Mora}}, \
  and\ \bibinfo {author} {\bibfnamefont {V.}~\bibnamefont {Hakim}},\
  }\href@noop {} {\bibfield  {journal} {\bibinfo  {journal} {arXiv preprint
  arXiv:1302.4424}\ } (\bibinfo {year} {2013})}\BibitemShut {NoStop}%
\bibitem [{\citenamefont {Jaynes}(1982)}]{jaynes1982rationale}%
  \BibitemOpen
  \bibfield  {author} {\bibinfo {author} {\bibfnamefont {E.~T.}\ \bibnamefont
  {Jaynes}},\ }\href@noop {} {\bibfield  {journal} {\bibinfo  {journal}
  {Proceedings of the IEEE}\ }\textbf {\bibinfo {volume} {70}},\ \bibinfo
  {pages} {939} (\bibinfo {year} {1982})}\BibitemShut {NoStop}%
\bibitem [{\citenamefont {Press{\'e}}\ \emph {et~al.}(2013)\citenamefont
  {Press{\'e}}, \citenamefont {Ghosh}, \citenamefont {Lee},\ and\ \citenamefont
  {Dill}}]{presse2013principles}%
  \BibitemOpen
  \bibfield  {author} {\bibinfo {author} {\bibfnamefont {S.}~\bibnamefont
  {Press{\'e}}}, \bibinfo {author} {\bibfnamefont {K.}~\bibnamefont {Ghosh}},
  \bibinfo {author} {\bibfnamefont {J.}~\bibnamefont {Lee}}, \ and\ \bibinfo
  {author} {\bibfnamefont {K.~A.}\ \bibnamefont {Dill}},\ }\href@noop {}
  {\bibfield  {journal} {\bibinfo  {journal} {Reviews of Modern Physics}\
  }\textbf {\bibinfo {volume} {85}},\ \bibinfo {pages} {1115} (\bibinfo {year}
  {2013})}\BibitemShut {NoStop}%
\bibitem [{\citenamefont {Roudi}\ \emph
  {et~al.}(2009{\natexlab{b}})\citenamefont {Roudi}, \citenamefont {Aurell},\
  and\ \citenamefont {Hertz}}]{roudi2009statistical}%
  \BibitemOpen
  \bibfield  {author} {\bibinfo {author} {\bibfnamefont {Y.}~\bibnamefont
  {Roudi}}, \bibinfo {author} {\bibfnamefont {E.}~\bibnamefont {Aurell}}, \
  and\ \bibinfo {author} {\bibfnamefont {J.~A.}\ \bibnamefont {Hertz}},\
  }\href@noop {} {\bibfield  {journal} {\bibinfo  {journal} {Frontiers in
  computational neuroscience}\ }\textbf {\bibinfo {volume} {3}} (\bibinfo
  {year} {2009}{\natexlab{b}})}\BibitemShut {NoStop}%
\bibitem [{\citenamefont {Kappen}\ and\ \citenamefont
  {Rodriguez}(1998)}]{kappen1998efficient}%
  \BibitemOpen
  \bibfield  {author} {\bibinfo {author} {\bibfnamefont {H.~J.}\ \bibnamefont
  {Kappen}}\ and\ \bibinfo {author} {\bibfnamefont {F.}~\bibnamefont
  {Rodriguez}},\ }\href@noop {} {\bibfield  {journal} {\bibinfo  {journal}
  {Neural Computation}\ }\textbf {\bibinfo {volume} {10}},\ \bibinfo {pages}
  {1137} (\bibinfo {year} {1998})}\BibitemShut {NoStop}%
\bibitem [{\citenamefont {Tanaka}(1998)}]{tanaka1998mean}%
  \BibitemOpen
  \bibfield  {author} {\bibinfo {author} {\bibfnamefont {T.}~\bibnamefont
  {Tanaka}},\ }\href@noop {} {\bibfield  {journal} {\bibinfo  {journal}
  {Physical Review E}\ }\textbf {\bibinfo {volume} {58}},\ \bibinfo {pages}
  {2302} (\bibinfo {year} {1998})}\BibitemShut {NoStop}%
\bibitem [{\citenamefont {Sessak}\ and\ \citenamefont
  {Monasson}(2009)}]{sessak2009small}%
  \BibitemOpen
  \bibfield  {author} {\bibinfo {author} {\bibfnamefont {V.}~\bibnamefont
  {Sessak}}\ and\ \bibinfo {author} {\bibfnamefont {R.}~\bibnamefont
  {Monasson}},\ }\href@noop {} {\bibfield  {journal} {\bibinfo  {journal}
  {Journal of Physics A: Mathematical and Theoretical}\ }\textbf {\bibinfo
  {volume} {42}},\ \bibinfo {pages} {055001} (\bibinfo {year}
  {2009})}\BibitemShut {NoStop}%
\bibitem [{\citenamefont {Aurell}\ and\ \citenamefont
  {Ekeberg}(2012)}]{aurell2012inverse}%
  \BibitemOpen
  \bibfield  {author} {\bibinfo {author} {\bibfnamefont {E.}~\bibnamefont
  {Aurell}}\ and\ \bibinfo {author} {\bibfnamefont {M.}~\bibnamefont
  {Ekeberg}},\ }\href@noop {} {\bibfield  {journal} {\bibinfo  {journal}
  {Physical review letters}\ }\textbf {\bibinfo {volume} {108}},\ \bibinfo
  {pages} {090201} (\bibinfo {year} {2012})}\BibitemShut {NoStop}%
\bibitem [{\citenamefont {Nguyen}\ and\ \citenamefont
  {Berg}(2012{\natexlab{a}})}]{nguyen2012bethe}%
  \BibitemOpen
  \bibfield  {author} {\bibinfo {author} {\bibfnamefont {H.~C.}\ \bibnamefont
  {Nguyen}}\ and\ \bibinfo {author} {\bibfnamefont {J.}~\bibnamefont {Berg}},\
  }\href@noop {} {\bibfield  {journal} {\bibinfo  {journal} {Journal of
  Statistical Mechanics: Theory and Experiment}\ }\textbf {\bibinfo {volume}
  {2012}},\ \bibinfo {pages} {P03004} (\bibinfo {year}
  {2012}{\natexlab{a}})}\BibitemShut {NoStop}%
\bibitem [{\citenamefont {Mastromatteo}(2013)}]{mastromatteo2013beyond}%
  \BibitemOpen
  \bibfield  {author} {\bibinfo {author} {\bibfnamefont {I.}~\bibnamefont
  {Mastromatteo}},\ }\href@noop {} {\bibfield  {journal} {\bibinfo  {journal}
  {Journal of Statistical Physics}\ }\textbf {\bibinfo {volume} {150}},\
  \bibinfo {pages} {658} (\bibinfo {year} {2013})}\BibitemShut {NoStop}%
\bibitem [{\citenamefont {Huang}(2013)}]{huang2013sparse}%
  \BibitemOpen
  \bibfield  {author} {\bibinfo {author} {\bibfnamefont {H.}~\bibnamefont
  {Huang}},\ }\href@noop {} {\bibfield  {journal} {\bibinfo  {journal} {The
  European Physical Journal B}\ }\textbf {\bibinfo {volume} {86}},\ \bibinfo
  {pages} {1} (\bibinfo {year} {2013})}\BibitemShut {NoStop}%
\bibitem [{\citenamefont {Cocco}\ and\ \citenamefont
  {Monasson}(2012)}]{cocco2012adaptive}%
  \BibitemOpen
  \bibfield  {author} {\bibinfo {author} {\bibfnamefont {S.}~\bibnamefont
  {Cocco}}\ and\ \bibinfo {author} {\bibfnamefont {R.}~\bibnamefont
  {Monasson}},\ }\href@noop {} {\bibfield  {journal} {\bibinfo  {journal}
  {Journal of Statistical Physics}\ }\textbf {\bibinfo {volume} {147}},\
  \bibinfo {pages} {252} (\bibinfo {year} {2012})}\BibitemShut {NoStop}%
\bibitem [{\citenamefont {Nguyen}\ and\ \citenamefont
  {Berg}(2012{\natexlab{b}})}]{nguyen2012mean}%
  \BibitemOpen
  \bibfield  {author} {\bibinfo {author} {\bibfnamefont {H.~C.}\ \bibnamefont
  {Nguyen}}\ and\ \bibinfo {author} {\bibfnamefont {J.}~\bibnamefont {Berg}},\
  }\href@noop {} {\bibfield  {journal} {\bibinfo  {journal} {Physical review
  letters}\ }\textbf {\bibinfo {volume} {109}},\ \bibinfo {pages} {050602}
  (\bibinfo {year} {2012}{\natexlab{b}})}\BibitemShut {NoStop}%
\bibitem [{\citenamefont {Huang}\ and\ \citenamefont
  {Kabashima}(2013)}]{huang2013adaptive}%
  \BibitemOpen
  \bibfield  {author} {\bibinfo {author} {\bibfnamefont {H.}~\bibnamefont
  {Huang}}\ and\ \bibinfo {author} {\bibfnamefont {Y.}~\bibnamefont
  {Kabashima}},\ }\href@noop {} {\bibfield  {journal} {\bibinfo  {journal}
  {Physical Review E}\ }\textbf {\bibinfo {volume} {87}},\ \bibinfo {pages}
  {062129} (\bibinfo {year} {2013})}\BibitemShut {NoStop}%
\bibitem [{\citenamefont {Cocco}\ and\ \citenamefont
  {Monasson}(2011{\natexlab{b}})}]{cocco2011adaptive}%
  \BibitemOpen
  \bibfield  {author} {\bibinfo {author} {\bibfnamefont {S.}~\bibnamefont
  {Cocco}}\ and\ \bibinfo {author} {\bibfnamefont {R.}~\bibnamefont
  {Monasson}},\ }\href@noop {} {\bibfield  {journal} {\bibinfo  {journal}
  {Physical review letters}\ }\textbf {\bibinfo {volume} {106}},\ \bibinfo
  {pages} {090601} (\bibinfo {year} {2011}{\natexlab{b}})}\BibitemShut
  {NoStop}%
\bibitem [{\citenamefont {Huang}(2010)}]{huang2010reconstructing}%
  \BibitemOpen
  \bibfield  {author} {\bibinfo {author} {\bibfnamefont {H.}~\bibnamefont
  {Huang}},\ }\href@noop {} {\bibfield  {journal} {\bibinfo  {journal}
  {Physical Review E}\ }\textbf {\bibinfo {volume} {81}},\ \bibinfo {pages}
  {036104} (\bibinfo {year} {2010})}\BibitemShut {NoStop}%
\bibitem [{\citenamefont {Aurell}\ \emph {et~al.}(2010)\citenamefont {Aurell},
  \citenamefont {Ollion},\ and\ \citenamefont {Roudi}}]{aurell2010dynamics}%
  \BibitemOpen
  \bibfield  {author} {\bibinfo {author} {\bibfnamefont {E.}~\bibnamefont
  {Aurell}}, \bibinfo {author} {\bibfnamefont {C.}~\bibnamefont {Ollion}}, \
  and\ \bibinfo {author} {\bibfnamefont {Y.}~\bibnamefont {Roudi}},\
  }\href@noop {} {\bibfield  {journal} {\bibinfo  {journal} {The European
  Physical Journal B-Condensed Matter and Complex Systems}\ }\textbf {\bibinfo
  {volume} {77}},\ \bibinfo {pages} {587} (\bibinfo {year} {2010})}\BibitemShut
  {NoStop}%
\bibitem [{\citenamefont {Barton}\ \emph {et~al.}(2014)\citenamefont {Barton},
  \citenamefont {Cocco}, \citenamefont {De~Leonardis},\ and\ \citenamefont
  {Monasson}}]{barton2014large}%
  \BibitemOpen
  \bibfield  {author} {\bibinfo {author} {\bibfnamefont {J.}~\bibnamefont
  {Barton}}, \bibinfo {author} {\bibfnamefont {S.}~\bibnamefont {Cocco}},
  \bibinfo {author} {\bibfnamefont {E.}~\bibnamefont {De~Leonardis}}, \ and\
  \bibinfo {author} {\bibfnamefont {R.}~\bibnamefont {Monasson}},\ }\href@noop
  {} {\bibfield  {journal} {\bibinfo  {journal} {arXiv preprint
  arXiv:1405.0233}\ } (\bibinfo {year} {2014})}\BibitemShut {NoStop}%
\bibitem [{\citenamefont {Hopfield}(1982)}]{hopfield1982neural}%
  \BibitemOpen
  \bibfield  {author} {\bibinfo {author} {\bibfnamefont {J.~J.}\ \bibnamefont
  {Hopfield}},\ }\href@noop {} {\bibfield  {journal} {\bibinfo  {journal}
  {Proceedings of the national academy of sciences}\ }\textbf {\bibinfo
  {volume} {79}},\ \bibinfo {pages} {2554} (\bibinfo {year}
  {1982})}\BibitemShut {NoStop}%
\bibitem [{\citenamefont {Cocco}\ \emph
  {et~al.}(2013{\natexlab{b}})\citenamefont {Cocco}, \citenamefont {Monasson},\
  and\ \citenamefont {Weigt}}]{cocco2013inference}%
  \BibitemOpen
  \bibfield  {author} {\bibinfo {author} {\bibfnamefont {S.}~\bibnamefont
  {Cocco}}, \bibinfo {author} {\bibfnamefont {R.}~\bibnamefont {Monasson}}, \
  and\ \bibinfo {author} {\bibfnamefont {M.}~\bibnamefont {Weigt}},\ }in\
  \href@noop {} {\emph {\bibinfo {booktitle} {Journal of Physics: Conference
  Series}}},\ Vol.\ \bibinfo {volume} {473}\ (\bibinfo {organization} {IOP
  Publishing},\ \bibinfo {year} {2013})\ p.\ \bibinfo {pages}
  {012010}\BibitemShut {NoStop}%
\bibitem [{\citenamefont {Tang}\ and\ \citenamefont
  {Sutskever}(2011)}]{tang2011data}%
  \BibitemOpen
  \bibfield  {author} {\bibinfo {author} {\bibfnamefont {Y.}~\bibnamefont
  {Tang}}\ and\ \bibinfo {author} {\bibfnamefont {I.}~\bibnamefont
  {Sutskever}},\ }\href@noop {} {\emph {\bibinfo {title} {Data normalization in
  the learning of restricted Boltzmann machines}}},\ \bibinfo {type} {Tech.
  Rep.}\ (\bibinfo  {institution} {Technical Report UTML-TR-11-2, Department of
  Computer Science, University of Toronto},\ \bibinfo {year}
  {2011})\BibitemShut {NoStop}%
\bibitem [{\citenamefont {Ciresan}\ \emph {et~al.}(2012)\citenamefont
  {Ciresan}, \citenamefont {Meier},\ and\ \citenamefont
  {Schmidhuber}}]{ciresan2012multi}%
  \BibitemOpen
  \bibfield  {author} {\bibinfo {author} {\bibfnamefont {D.}~\bibnamefont
  {Ciresan}}, \bibinfo {author} {\bibfnamefont {U.}~\bibnamefont {Meier}}, \
  and\ \bibinfo {author} {\bibfnamefont {J.}~\bibnamefont {Schmidhuber}},\ }in\
  \href@noop {} {\emph {\bibinfo {booktitle} {Computer Vision and Pattern
  Recognition (CVPR), 2012 IEEE Conference on}}}\ (\bibinfo {organization}
  {IEEE},\ \bibinfo {year} {2012})\ pp.\ \bibinfo {pages}
  {3642--3649}\BibitemShut {NoStop}%
\bibitem [{\citenamefont {Salakhutdinov}\ \emph {et~al.}(2007)\citenamefont
  {Salakhutdinov}, \citenamefont {Mnih},\ and\ \citenamefont
  {Hinton}}]{salakhutdinov2007restricted}%
  \BibitemOpen
  \bibfield  {author} {\bibinfo {author} {\bibfnamefont {R.}~\bibnamefont
  {Salakhutdinov}}, \bibinfo {author} {\bibfnamefont {A.}~\bibnamefont {Mnih}},
  \ and\ \bibinfo {author} {\bibfnamefont {G.}~\bibnamefont {Hinton}},\ }in\
  \href@noop {} {\emph {\bibinfo {booktitle} {Proceedings of the 24th
  international conference on Machine learning}}}\ (\bibinfo {organization}
  {ACM},\ \bibinfo {year} {2007})\ pp.\ \bibinfo {pages} {791--798}\BibitemShut
  {NoStop}%
\end{thebibliography}%

\end{document}